\documentstyle[15lomcon,cite]{article}

\RequirePackage{xspace}
\def\superb{Super$B$\xspace}

\def\lhc {LHC\xspace}

\def\babar{\mbox{\slshape B\kern-0.1em{\smaller A}\kern-0.1em
    B\kern-0.1em{\smaller A\kern-0.2em R}}\xspace}

\begin{document}

\title{THE PHYSICS PROGRAMME AT SUPER$B$}

\author{ Adrian Bevan (on the behalf of the \superb Collaboration) \footnote{e-mail: a.j.bevan@qmul.ac.uk}
}

\address{ Department of Physics, Queen Mary, University of London, Mile End Road, London, E1 4NS, UK }

\maketitle\abstracts{ \superb is a next generation high luminosity $e^+e^-$ collider
that will be built at the Cabibbo Laboratory, Tor Vergata, in Italy.  The physics goals
of this experiment are to search for signs of physics beyond the Standard Model through 
precision studies of rare or forbidden processes.  While the name suggests that $B$ physics
is the main goal, this experiment is a Super Flavour Factory, and precision measurements
of $B_{u,d,s}$, $D$, $\tau$, $\Upsilon$, and $\psi(3770)$ decays as well as spectroscopy 
and exotica searches form part of a broad
physics programme.  In addition to searching for new physics (NP) in the form of heavy particles,
or violations of laws of physics, data from \superb will be able to perform
precision tests of the Standard Model.
I will briefly review of some highlights of the \superb 
physics programme.}

\section{Introduction}

Flavour physics has been instrumental in the development of the Standard Model 
(SM), starting from a combination of results from hyperon decays
by Cabibbo~\cite{Cabibbo:1963yz}.  Shortly after this development Glashow, Iliopolus 
and Maiani~\cite{Glashow:1970gm} proposed the existence of the charm quark to 
satisfy the observed pattern of branching fractions in kaon decays and establish 
a four quark model of particle physics.  The discovery of CP violation by 
Cronin, Christenson, Turlay and Fitch in 1964~\cite{Christenson:1964fg} was another 
landmark, and in time Kobayashi and Maskawa postulated 
a six quark model in order to accommodate CP violation naturally within the 
SM~\cite{Kobayashi:1973fv}.  These developments laid the foundations of our current theory, 
however we know that there are a number of features missing, including
an understanding of the matter-antimatter asymmetry.

\superb is a next generation $e^+e^-$ flavour factory designed to operate 
primarily with a centre of mass energy of the $\Upsilon(4S)$, and at 
the charm threshold $\psi(3770)$.  The project status, accelerator and 
detector are discussed in the contribution to these proceedings by Francesco 
Forti~\cite{forti}.  Recent reviews of the physics programme of this project
can be found in Refs.~\cite{interplay,physicswp,review}.  The rest of these proceedings
discuss a number of physics highlights, and how these can be used to 
elucidate the structure of physics beyond the SM.

%
% tau Physics
%
\section{$\tau$ Physics}
The intrinsic level of charged lepton flavour violation in the $\tau$ sector arising from 
neutrino oscillations is expected to occur at the level of $10^{-54}$.  Given that both quark 
and neutral lepton number conservation is known to be violated at a small level, it is natural
to presume that there may ultimately be non-conservation of charged lepton number.  Indeed
many scenarios of physics beyond the SM predict large (up to $\sim 10^{-9}$)
levels of charged lepton flavour violation (LFV).  These predictions are model dependent: some
models favour large $\mu \to e$ transitions over other possibilities.  Other models prefer
large $\tau\to \mu$ or $\tau\to e$ transitions.  While the quest for a discovery of LFV
continues, it is clear that all three sets of transitions need to be measured or well constrained
in order to understand the underlying dynamics.  \superb will be able to improve upon existing
limits from the $B$ factories by between one and two orders of magnitude.  Channels such as
$\tau \to \ell \gamma$ will see a factor of ten improvement as these have irreducible 
SM backgrounds that one will have to contend with, while other channels such as $\tau^\pm \to \ell^+\ell^-\ell^\pm$,
which are free of SM backgrounds, will see a factor of one hundred improvement.

The $e^-$ beam at \superb will be 80\% polarised, enabling one to separate 
contributions from SM-like LFV channels and otherwise irreducible backgrounds as one
can use the polarisation of the final state $\tau$ lepton produced in collisions in 
order to suppress background.  This works well for improving limits, or indeed searching 
for left handed sources of NP.  One can verify if there is a right handed component
to any underlying NP by comparing results with and without polarised beams.
Similarly as one expects some higher theory to undergo symmetry breaking in order to manifest
the low energy scenarios we are studying at facilities today, which includes the 
LHC reach, the different types of fundamental particle may be related to each other.
There are models that predict correlated effects between charged leptons decays and
processes involving quarks, and of course neutrinos.  Hence in order to understand 
any underlying deviations in any of these sectors one needs to cross check, for example,
charged LFV results with neutrino mixing, neutral meson mixing and CP violation measurements
in the quark sector.  A number of scenarios of physics beyond the SM are discussed in Ref.~\cite{physicswp}
in order to illustrate this point.

%
% B Physics
%
\section{$B$ Physics}
As much of the NP search potential of \superb concerns the use of 
indirect constraints on rare processes to infer the existence or otherwise
of some model, one naturally has some sensitivity to the corresponding 
energy scale $\Lambda_{NP}$ of the NP.  The correlation between 
measurement of the rare decays (a branching fraction or other observable)
and the energy scale is non trivial.  If one considers the minimal supersymmetric 
model (MSSM) in the mass insertion hypothesis then for example
the measurement of
the inclusive branching fractions of $b\to s\gamma$ and $b\to s \ell \ell$,
along with the $CP$ asymmetry in $b\to s\gamma$ can be used to constrain
the mass insertion parameter $(\delta^d_{23})_{LR}$.   The magnitude of this
parameter can be used to infer an upper limit on $\Lambda_{NP}$ to complement 
the null results obtained so far from the LHC.  If generic MSSM was a realistic
description of nature then the fact that the LHC has failed to find a low
mass gluino implies that there is a non-trivial coupling $(\delta^d_{23})_{LR}$, and
hence in turn \superb should be able to observe a non-trivial deviation
from the SM when studying the inclusive decays $b\to s\gamma$ and $b\to s \ell \ell$.
The magnitude of the observed deviation will benefit the SLHC community as
the inferred upper bound on the energy scale obtained will provide
useful information on the integrated luminosity required to yield positive
results via direct searches.  For example if one measured $|(\delta^d_{23})_{LR}| = 0.05$,
then the implied upper limit on $\Lambda_{NP}$ is 3.5 TeV, which is also
compatible with known constraints on $\tan\beta$ as can be seen from Ref.~\cite{physicswp}.
Other processes can be used to constrain other mass insertion parameters.

There are a number of golden rare $B$ decay channels at \superb, including 
$B\to \ell \nu$, where $\ell = \tau, \mu, e$.  In the SM this decay 
is known up to uncertainties relating to the value of $V_{\mathrm{ub}}$ and
$f_B$.  The rate of these processes can be modified by the existence of 
charged Higgs particles predicted in a number of extensions of the SM for
example two-Higgs Doublet models (2HDM) or SUSY extensions of the SM.
Hence the measured rate of these decays can be used to place limits on the
inferred mass of any $H^+$ particle, and such constraints are a function
of $\tan\beta$.  Existing
constraints from the $B$ factories from inclusive $b\to s\gamma$ decays
exclude masses below $295 GeV/c^2$, and the constraint from $B\to \tau \nu$
excludes higher masses for large $\tan\beta$ scenarios.  With $75ab^{-1}$
of data \superb will be able to exclude, or detect, a $H^+$ with 
a mass $1-3$ TeV, for $\tan\beta$ between 40 and 100.  This constraint 
results from a combination of $B\to \tau \nu$ (dominates at lower
luminosity) and $B\to \mu \nu$ (dominates at high luminosity).  Ref.~\cite{physicswp}
discusses the physics potential of a number of other interesting rare $B$ decays.

Many of the $CP$ asymmetry observables of $B_{u,d}$ decays available at \superb 
are dominated by loop contributions and are sensitive to the 
same sources of NP that can affect many of the interesting rare decays
discussed above.  The golden modes to use
in measuring the angle $\beta$ of the unitarity triangle are $B$
decays to charmonium ($c\overline{c}$), $\eta^\prime$ or $\phi$ and a neutral kaon.
\superb will be able to measure the $CP$ asymmetries in these decays with 
precisions of 0.002, 0.008, and 0.021, respectively using a data sample
of $75ab^{-1}$.  Both tree ($c\overline{c}K^0$) and penguin dominated decays
can be affected by the presence of NP.

To complement the $B_{u,d}$ programme at \superb, there will be a dedicated
run at the $\Upsilon(5S)$ resonance which enables the study of a number of 
$B_s$ related observables that may be affected by physics beyond the SM.  
These include the semi-leptonic asymmetry and branching fraction $B_s\to \gamma\gamma$.

%
% D Physics
%
\section{Charm Physics}
Charm mixing has been established by the $B$ factories and is parameterised
by two small numbers: $x=\Delta m_{D} / \Gamma$ and $y=\Delta \Gamma / 2\Gamma$.
These are currently measured as $x= (0.65^{+0.18}_{-0.19})\%$, and $y= (0.74\pm 0.12)\%$~\cite{hfag}.
The precision with which these mixing parameters can be improved upon is dominated
by inputs from $D^0 \to K_S h^+h^-$ decays, where $h=\pi$, $K$.  At large integrated
luminosities one of the limiting factors of this analysis will be the knowledge of
the strong phase variation across the $K_S h^+h^-$ Dalitz plot.  This can be 
measured using data collected at charm threshold, where $e^+e^-\to \psi(3770)\to D^0\overline{D}^0$
transitions result in pairs of quantum correlated neutral $D$ mesons.  These
correlated mesons can be used to precisely determine the required map of 
the strong phase difference required for charm mixing measurements.  With this
input from a data sample of $500 fb^{-1}$ \footnote{\superb is expected to accumulate twice
this luminosity at charm threshold} the mixing measurements at \superb will
still be statistics limited, and one should be able to achieve precisions 
of $0.02\%$ and $0.01\%$ on $x$ and $y$, respectively.  The strong phase difference
map measured at charm threshold will also be an important input used for the 
determination of the unitarity triangle angle $\gamma$ for \superb, Belle II
and LHCb.

It has often been said that $CP$ violation in charm will provide a unique test
of the SM. The reason for this is that one expects only very small effects in the 
SM, and so any large measured deviation from zero would be a clear sign of 
NP.  Just like the $B_{u,d}$ system, charm has a unitarity triangle 
that needs to be tested.  The physics potential of \superb in this area 
has recently been outlined in Ref.~\cite{Bevan:2011up}.  In the months
following the Lomonosov conference an intriguing hint of $CP$ violation
in charm decays was produced by the LHCb experiment~\cite{LHCbconf}.  This relates to
a difference in direct $CP$ asymmetry parameters measured in $D\to KK$
and $D\to \pi\pi$.  If this is a real effect one will have to perform
the measurements outlined in~\cite{Bevan:2011up} in order to understand
the underlying physics.

A number of charm rare decay analogues of the $B$ physics programme are of interest in
constraining NP and elucidating the underlying decay dynamics in charm.
The complication with charm is the presence of long distance
dynamics that dominate the final states.  The combination of charm threshold
and $\Upsilon(4S)$ running naturally complement each other when studying rare
charm decays.

\section{Precision Electroweak Physics}
In terms of precision electroweak physics, the polarised electron beam at \superb
facilitates the measurement of $\sin^2\theta_W$ at centre of mass energy corresponding
to the $\Upsilon(4S)$ resonance. \superb will be able to measure this quantity with
the same precision as the LEP/SLC measurements made at the $Z^0$ pole.  However 
the advantage of the \superb measurement is the fact that the $e^+e^- \to b\overline{b}$
result will be free from fragmentation uncertainties that limit the interpretation of 
the LEP/SLC measurements. The \superb measurements will be complementary to other
low energy $\sin^2\theta_W$ measurements from the QWeak Collaboration (JLab), and at the 
proposed MESA experiment (Mainz).

\section{Interplay Between Measurements and Summary}

The power of \superb comes from the ability to study a diverse set of modes
that are sensitive to different types of NP.  Through the 
pattern of deviations from SM expectations for 
observables one will be able to identify viable NP scenarios and reject
those that are not compatible with the data.  This goes beyond the motivation
of simply discovering some sign of NP and is a step toward developing
a detailed understanding of NP.  If no significant deviations are 
uncovered then this in turn can be used to constrain parameter space and 
reject models that are no longer viable.  Given that many of the observables
that \superb will measure are not accessible directly at the \lhc, these results
will complement the direct and indirect searches being performed at CERN.
Detailed discussions on the interplay problem can be found in 
Refs.~\cite{interplay,physicswp}.

%\section{Summary}

In summary the physics programme at \superb is varied, and the unique features of
the facility: polarised electron beams and a dedicated charm threshold run add to 
its strengths via versatility.
The charm threshold run in particular, in addition
to facilitating a number of NP searches, will provide several measurements
required to control systematic uncertainties for measurements of 
charm mixing and the unitarity triangle angle $\gamma$.  Results from \superb will
be able to play a role in elucidating any NP discovered at the \lhc and
indirectly probe to higher energy than the \lhc will be able to directly
access.


\section*{References}
\begin{thebibliography}{99}

\bibitem{Cabibbo:1963yz}
N. Cabibbo, Phys. Rev. Lett. {\bf 10}, 531-533 (1963).

\bibitem{Glashow:1970gm}
S. Glashow, J. Iliopoulos, and L. Maiani, Phys. Rev. D{\bf 2}, 1285-1292 (1970).

\bibitem{Christenson:1964fg}
H. Christenson {et et al.}, Phys. Rev. Lett. {\bf 13}, 138-140 (1964).

\bibitem{Kobayashi:1973fv}
M. Kobayashi and T. Maskawa, Prog. Theor. Phys. {\bf 49}, 652-657 (1973).

\bibitem{forti} F. Forti, in {\it ``Particle Physics on the Eve of LHC''}
               (These proceedings of the 15th Lomonosov
               Conference on Elementary Particle Physics, August 18-24 2011,
               Moscow, Russia), ed. by A.Studenikin.

\bibitem{interplay} B. Meadows {et et al.}, arXiv:1109.5028.

\bibitem{physicswp} B. O'Leary {et et al.}, arXiv:1008.1541.

\bibitem{review} A. Bevan, arXiv:1110.3901.

\bibitem{hfag} 
Heavy Flavour Averaging Group (HFAG), {\tt http://www.slac.stanford.edu/xorg/hfag/}.

\bibitem{Bevan:2011up}
A. Bevan, G. Inguglia, and B. Meadows, arXiv:1106.5075.

\bibitem{LHCbconf}
LHCb Collaboration, LHCb-CONF-2011-023.

\end{thebibliography}
\end{document}